\documentclass[12pt]{article}
\usepackage{amssymb,amsmath,amsthm}
\usepackage{rotating}
\usepackage{graphicx,comment}
\usepackage{color}

\begin{document}


\begin{center}

{\bf On the strong instability of the  multi-layer Hele-Shaw  flows.}

Gelu I. Pa\c sa,  Simion Stoilow Institute  of Mathematics   
of Romanian Academy 

e-mail:   gelu.pasa@imar.ro

\end{center}

We study  the effects of some  injection policies used  in oil recovery process. 
The Saffman-Taylor instability occurs when a less viscous fluid is displacing a  more  
viscous one, in a rectangular Hele-Shaw cell. 
The injection of $N$   successive intermediate phases  with constant viscosities (the 
multi-layer Hele-Shaw model) was studied in some recent papers, where a minimization
of the Saffman-Taylor instability was obtained for large enough $N$.
However, in  this  paper we get a  particular  eigenfunction of the linear stability 
system which leads to  eigenvalues which become infinite  for large wave numbers. 
We obtain  a strong instability of the multi-layer  Hele-Shaw displacement,  even if 
$N$  is very large.

\vspace{0.5cm}



\begin{center} 

{\bf   1. Introduction.} 

\end{center}

We consider  a  Stokes flow in a Hele-Shaw cell (see \cite{HS}) parallel  
with the plane $xOy$. 
The  thickness of the gap between the cell plates is denoted by $b$. The 
gravity  effects are neglected. 
The  viscosity,  velocity  and  pressure are denoted by  $\nu,  {\bf u}=
(u,v,w), p$. As $b$ is very small, we neglect $w$. The flow equations are 
$$
 p_x = -  \frac{12 \nu }{b^2} <u>, \quad 
 p_y = -  \frac{12 \nu }{b^2} <v>, \quad p_z=0,                         $$
\begin{equation}\label{HS-4}
   <u>_x+<v>_y=0,                                                          
\end{equation}
where the lower indices  $x,y,z$ are denoting the partial derivatives and
  $< F > =(1/b) \int_0^b F dz $.
The above equations are similar to the  Darcy's law for the flow in  a porous 
medium  with the  permeability  $(b^2/12) $ - see  \cite{BE}, \cite{LAMB}.

A sharp interface exists between two immiscible  displacing fluids in a Hele -
Shaw cell.  
This flow-model can be used  to study the secondary oil-recovery process: the 
oil (with low pressure) contained in a porous medium is obtained by pushing 
it with a second  displacing fluid. Saffman  and Taylor \cite{SAFF-TAY}  have 
proven  the well know  result: 
the interface is unstable  when the displacing fluid  is less viscous. Moreover,  
the fingering phenomenon appears in this case - see   \cite{HOM}, \cite{SAFF}. 

The optimization of  displacements in porous media were studied in 
\cite{CHEN}, \cite{DIAZ}, \cite{AL-HUS-1},  \cite{AL-HUS-2}, \cite{SUDARYANTO}.
Some effects due to the very small surface tensions are studied in \cite{TA1}, 
\cite{TA2}.

An intermediate fluid  with  a variable  viscosity in a middle layer between 
the  displacing fluids can minimize  the  Saffman-Taylor instability - see  the 
experimental  and numerical results  given in \cite{GILJE},  \cite{GOR-HOM-1}, 
\cite{GOR-HOM-2}, \cite{MUN}, \cite{SHAH}, \cite{SLOBOD}, \cite{UZOIGWE}.

A theoretical optimal  variable intermediate viscosity was obtained in  Carasso and 
Pasa \cite{CARASSO}.  An important improvement of stability of the  was proved, 
compared with the Saffman-Taylor case. 

A linear stability  analysis of the three-layer  Hele-Shaw flow with a constant 
intermediate viscosity was performed in \cite{DAR-1}, by using an algebraic
method.

The Hele-Shaw displacement with $N$ intermediate layers was studied in \cite{DAR-2}, 
\cite{DAR-DING}, \cite{D7}, \cite{D6}. We use the notation MLHS  for this Multi-Layer 
Hele-Shaw model. 
In the case of constant intermediate viscosities with  positive  jumps in the flow 
direction,  an improvement of the stability was obtained in \cite{DAR-2},  
\cite{DAR-DING}, \cite{D7}, \cite{D6}, if the  number  of the intermediate  layers is
 very  large  and  the  surface tensions  verify   some  conditions.

In the present paper we obtain  a new   eigenfunction of the linear  system which 
governs the linear stability  of MLHS. The corresponding  eigenvalues become  infinite 
for large wave numbers. Our result is mainly  builds upon the  boundary conditions on 
the interfaces where the  viscosity jumps exist. Then we get  a strong instability
of MLHS flow, even if $N$ is very large.   
Moreover, we prove that large values of the surface tensions on the  interfaces are 
amplifying the  instability, in contradiction with the experimental results and with the 
Saffman-Taylor  criterion.  

In the last part of this paper, we show that a possible  strategy to minimize the Saffman-Taylor 
instability is the use of an  intermediate fluid with a suitable {\it variable}  viscosity.

The paper is laid out as follows. 
In section 2   we recall the three-layer Hele-Shaw model introduced in \cite{GOR-HOM-1}. 
In section 3 we use this result  for a model with  $N$ intermediate layers with constant 
viscosities and we prove the flow instability.
In section 4 we study the effect of an intermediar  linear  viscosity profile. We conclude 
in section 5.

\newpage

\begin{center}

{\bf  2.  The three-layer Hele-Shaw model.} 

\end{center}

The three-layer Hele-Shaw flow with variable  intermediate viscosity was first described in 
\cite{GOR-HOM-1}   and studied also in \cite{GOR-HOM-2}. We recall here the  basic elements.  

A polymer solute with a variable concentration $c$ and variable viscosity $\nu$ is  injected  
with the positive velocity $U$ in a rectangular Hele-Shaw cell which is saturated with oil of 
viscosity  $ \nu_O$,  during a time interval $TI$.  
As in \cite{GOR-HOM-1},  adsorption, dispersion and diffusion of the solute in the equivalent 
porous medium are neglected. The  expression of the intermediate viscosity  $\nu$ as a function 
of $c$  is
\begin{equation}\label{ZT02}
 \nu(c)  = a_0 + a_1 c + a_2c^2 + .... 
\end{equation}
where $a_i$ are  constant coefficients  - see  \cite{FLORY},  \cite{GILJE}. 
In the case  of a dilute  solute,  which is studied here, we have  $ \nu = a_0 + a_1 c $,  then 
$ \nu $ is invertible  with respect to  $c$.
The continuity equation for the solute is $Dc/Dt=0$, then we have $D \nu / Dt=0$. That means 
\begin{equation}\label{ZT003}
\nu _t + u \nu_x  + v \nu_y = 0.
\end{equation}
After $TI$,  a displacing fluid   with viscosity $\mu_W$ is injected in the porous medium,  with  
the  same velocity $U$. 

We consider  incompressible fluids, then the amount of the polymer solute between the two interfaces
cannot change, according to the  principle   of mass   conservation. Therefore an arbitrary (small) 
movement  of  the first  interface must induce a movement with the same velocity  of the second 
 interface.

However, it is well known - see Saffman and Taylor \cite{SAFF-TAY} - that interfaces  change over time 
and turn into fingers of fluid (or polymer solute).  
We study the  evolution  of perturbations only  in a small time interval after $TI$ and  believe that 
the  initial  shape  of interfaces has not changed so  much.
On this way we  obtain   an  intermediate  fluid layer, moving with the velocity $U$, where the 
viscosity is variable. 

Consider $u=U, v=0$, then from   \eqref{ZT003}   we get $  \nu = \nu(x-Ut)$.
In some  experiments  (see  \cite{MUN})  an exponentially - decreasing (from the front  interface) 
viscosity   $\nu(x-Ut)$  was used  and  the instability  was almost   suppressed. The displacements 
with variable   viscosity in Hele-Shaw cells   and   porous media  are  studied in \cite{CARASSO},
 \cite{LOGGIA},  \cite{TALON}.

The displacing  and the  displaced fluids are denoted with the lower indices $_W, \,\, _O$.

Suppose the intermediate region  is the interval 
$$ Ut < x <  Ut + L,                                                                              $$ 
moving with the  constant   velocity $U$ far upstream. We have three incompressible fluids with the
viscosities  $\nu_W$ (displacing fluid),  $ \nu $ (intermediate  layer) and  $ \nu_O$ (displaced 
fluid). The flow is governed by    the Darcy's  equations, where $(u,v)$ are the averaged velocities 
- see \eqref{HS-4}:  
\begin{equation}\label{ZT004}
  p_{x} = - \mu_d u; \quad  p_y = -\mu_d v;  
 \quad p_z=0;    \quad 
 u_{x} + v_y = 0;                                                           
\end{equation}
$$
\mu_d = \mu_W, \,\, x < Ut;          \quad
   \mu_d = \mu,   \,\, x \in ( Ut, \,\,  Ut + L ) ;  
\quad    \mu_d = \mu_O, \,\, x > Ut + L ;       $$
\begin{equation}\label{DIM-VISCO}
\mu_W = 12\nu_W /b^2; \quad  \mu = 12 \nu /b^2; 
\quad   \mu_O =  12 \nu_O/b^2.                                                                  
\end{equation}

We consider the following basic state.  The  basic velocity 
and interfaces are 
$$ u=U, \,\, v=0; \quad  \quad  x = Ut, \,\, x =Ut+L. $$ 

On the interfaces we consider the  Laplace's law: the 
pressure jump is given by the surface tension multiplied with 
the interfaces curvature and the  component $u$ of the velocity 
is continuous.  Moreover,  the interface is a material one. The 
basic interfaces are straight lines,  then the basic
 pressure $P$ is  continuous (but his gradient is not) and          
 \begin{equation}\label{BASIC-PRESS}
P_{x}= -  \mu_d U, \quad P_y=0.
\end{equation}
We use the equation  \eqref{ZT003}, then the basic (unknown) 
viscosity $\mu$ in the middle layer verifies  the equation
\begin{equation}\label{ZT004A}
 \mu_t + U \mu_{x} = 0.
\end{equation}
We introduce the moving reference frame 
\begin{equation}\label{MR}
{\overline x}=x - Ut, \quad    \tau = t.                
\end{equation}
The equation  \eqref{ZT004A}  leads to $\mu_{\tau}=0$, then  
$\mu = \mu ({\overline x}) $. The intermediate region in the  
moving  reference frame is the segment $0 <{\overline x} < L$. 
We   still use  the notation  $x, \,\, t$ instead of 
${\overline x}, \tau$.

\vspace{0.25cm}

The small perturbations   of the basic velocity, pressure and viscosity 
are denoted by $u', v', p', \mu'$. We  insert the perturbations  in the  
equations \eqref{ZT004}, \eqref{ZT004A}. As in \cite{GOR-HOM-1}, we obtain  
the  linear stability system:
\begin{equation}\label{ZT005A}
p'_x = -\mu u' - \mu' U, \quad p'_y = -  \mu v', \quad 
 u'_x + v'_y = 0,
\end{equation}
\begin{equation}\label{ZT006A}
 \mu'_t + u' \mu_x = 0. 
\end{equation}

The above system is linear in disturbance quantities. Consider a   perturbation 
velocity $u'$ of the form:
\begin{equation}\label{FOURIER-U}
 u'(x,y,t) =                                                       
f(x)  [   \cos(ky) + \sin (ky)] e^{\sigma t}, \,\, k \geq 0,  
\end{equation}
where   $f(x)$ is the amplitude, $\sigma $ is the growth constant  and  $k$ are 
the wave numbers.

The  velocity along the axis $Ox$ is continuous, then the amplitude 
$f(x)$ is  continuous. From   $\eqref{ZT005A}_3$,  $\eqref{ZT005A}_2$, 
\eqref{ZT006A},  \eqref{FOURIER-U}  we get 
$$ v' = ( 1/k) f_x 
[-  \sin(ky) + \cos(ky)] e^{\sigma t},                                      $$
$$ p' = (\mu / k^2) f_x 
[-  \cos(ky) - \sin(ky)] e^{\sigma t},                                      $$                               
\begin{equation}\label{ZT007}
 \mu' = (-1/ \sigma) \mu_x f
[  \cos(ky) +  \sin (ky)]e^{\sigma t}.           
\end{equation} 
We perform the cross derivation of the relations $\eqref{ZT005A}_1, 
\eqref{ZT005A}_2$ and obtain
$$ \mu u'_y + \mu'_y U =  \mu_x v' + \mu v'_{x} .              $$
Then from  $\eqref{FOURIER-U}, \eqref{ZT007}_1,  \eqref{ZT007}_3$ 
we get the  equation which  governs   the amplitude  $f$:
\begin{equation}\label{ZT008}
 -(\mu f_x)_x +  k^2 \mu f = \frac{1}{\sigma} U k^2 f \mu_x,  
 \quad \forall x \notin \{0,L \}.
\end{equation}

The viscosity is constant outside the intermediate region, then  
from \eqref{ZT008}  we get 
$$ -f_{xx} + k^2 f = 0, \quad x \notin (0,L)    .              $$
The perturbations must decay to zero in   the far field. Then  we
have
\begin{equation}\label{FAR-FIELD}
f(x) = f(0) e^{ kx }, \,\,       x \leq 0; \quad 
f(x) = f(L) e^{ -k(x-L) }, \,\,  x \geq L.
\end{equation}

We  describe the Laplace law in a point $x=a$ where a  a viscosity 
jump exists.
The amplitude $f$ is continuous in $a$ but $f_x$ could be discontinuous. 
The perturbed interface near $a$ is denoted by $\eta'(a,y,t)$. In the 
first approximation we have  $\eta'_t = u'$. As in \cite{GOR-HOM-1}, we 
consider
\begin{equation}\label{INTER001}
\eta'(a,y,t) =                                                   
(1/ \sigma_a)  f(a)  [  \cos(k y) +  \sin(ky) ] e^{\sigma t },
\end{equation}  
where $\sigma_a$ is the growth rate associated to the  point $x=a$.

We search  for the right and left limit values  of the pressure in 
the point   $a$, denoted  by $p^+(a), \quad p^-(a)$. 
For this we use  the basic pressure $P$   in the point  $a$, the 
Taylor first order  expansion  of $P$ near $a$ and the expression   
$\eqref{ZT007}_2$ of  $p'$ in $x=a$. From  \eqref{BASIC-PRESS} it  
follows 
$$P_x^{+,-}(a)= -\mu^{+,-}(a)U                              $$   
then  we get 
$$
p^+(a) = P^+(a) + P^+_x(a) \eta + p'^+(a), \quad 
p^-(a) = P^-(a) + P^-_x(a) \eta + p'^-(a),                   $$
\begin{equation}\label{INTER002}
p^+(a)=  P^+(a) - \mu^+(a)  
\{ \frac{U f(a)}{\sigma_a} +\frac{ f_x^+(a)}{k^2} \}
[ \cos(k y) +  \sin(ky) ]e^{\sigma t},      
\end{equation}              
\begin{equation}\label{INTER003}
p^-(a) = P^-(a)  - \mu^-(a)  
\{ \frac{U f(a)}{\sigma_a} +\frac{ f_x^-(a)}{k^2} \}
[ \cos(k y) +  \sin(ky) ]e^{\sigma t}.       
\end{equation}               
The  Laplace's law is
\begin{equation}\label{LAPLACE001}
  p^+(a) - p^-(a) = T(a) \eta_{yy},
\end{equation}
where $T(a)$ is the surface tension acting in the point $a$ 
and  $\eta_{yy}$  is the approximate value  of the  curvature 
of  the perturbed  interface. As $P^-(a)=P^+(a)$,  from the 
equations \eqref{INTER002} - \eqref{LAPLACE001} we get  the 
relationship between    $f_x^-(a)$, $ f_x^+(a)$ and $\sigma$:
\begin{equation}\label{LAPLACE002}
- \mu^+(a)[\frac{Uf(a)}{\sigma_a}+  \frac{f_x^+(a)}{k^2}] +     
  \mu^-(a)[\frac{Uf(a)}{\sigma_a}+  \frac{f_x^-(a)}{k^2}] =        
-  \frac{T(a)}{\sigma_a}f(a)k^2.
\end{equation}

 When  $L=0$, from \eqref{FAR-FIELD} it follows 
$$f_x^-(0) =  k f(0),   \quad  f_x^+(0)=-k f(0), \quad 
  \mu^-(0) =  \mu_W,   \quad  \mu^+(0)= \mu_O.                $$ 
Then from \eqref{LAPLACE002} we recover   the Saffman - Taylor 
formula
\begin{equation}\label{SIGMA_ST01} 
\sigma_{ST} = \frac{ k U(\mu_O - \mu_W) - T(a) k^3}{\mu_O + \mu_W}.
\end{equation}
If $\mu_O > \mu_W$,  then $\sigma_{ST}>0$ in the range
$  k^2 <  U(\mu_O - \mu_W)/T(a)                               $
and the flow is unstable.

\begin{center}

 {\bf 3. The  $N$-layers Hele-Shaw displacements  with constant viscosities. } 

\end{center}

It is possible  to inject several  polymer-solutes  with constant-concentrations  
$c_1,c_2,..., c_N $  during the time intervals $TI_1, TI_2, ..., TI_N, \,\, N>1$. 
We  get a steady flow of $N$  layers of immiscible fluids  with {\it constant} 
viscosities $\mu_i$ such that (see the equation \eqref  {ZT008})
\begin{equation}\label{STRATELE-A}
 \mu_W = \mu_0 < \mu_1 < \mu_2< ... < \mu_N  <  \mu_{N+1}= \mu_O; 
\quad    \mu(x)=\mu_{i+1}, \,\, x\in (x_i, x_{i+1});                                           
\end{equation}
\begin{equation}\label{STRATELE-B}
  0= x_0 <x_1<x_2...<x_i<x_{i+1}<...< x_N=L;                                 
\end{equation}
 \begin{equation}\label{STRATELE-2}
-\mu_{i+1} f_{xx} + \mu_{i+1}k^2f =0,  \,\, x\in (x_i, x_{i+1}), \quad 
i=0,1,2,...,N-1.                                
\end{equation} 
The  viscosities jumps are   positive  in the flow direction. On each interface 
$x_i$ we have the boundary conditions \eqref{LAPLACE002}, which can  be written
in the form
\begin{equation}\label{STRATELE-3}    
\mu^-(x_i) f_x^-(x_i) - \mu^+(x_i) f_x^+(x_i)  = 
\frac{k^2 U[ \mu^+(x_i)- \mu^-(x_i)]-k^4T(x_i)}{\sigma_i}f(x_i),
\end{equation}
where $\sigma_i$ is the eigenvalue corresponding to the interface $x_i$.

This  $N$-layers model is studied in  \cite{DAR-2}, \cite{DAR-DING}, \cite{D7}, 
\cite{D6}. The boundary conditions  \eqref{STRATELE-3}  of our paper are equivalent
with the  equations (20) of \cite{DAR-2}.

In this section we show that the  $N$-layer Hele-Shaw  flow with constant 
intermediate viscosities is much more unstable, compared with  the Saffman-Taylor 
displacement. We get a  new particular eigenfunction of the linear stability  system 
which gives  us  positive eigenvalues which become infinite  for large wave numbers, 
even if $N$ and the surface tensions are very large.

\vspace{1.5cm}

{\it Propositin 1. } There exists at least a growth rate $\sigma_i$ 
corresponding to  \eqref{STRATELE-2}-\eqref{STRATELE-3}   s.t.  
$$ \sigma_i \rightarrow \infty \mbox{  for large } k.            $$
{\it Proof.} We consider the following  eigenfunctions which verify 
the relations \eqref {FAR-FIELD} and \eqref{STRATELE-2}:  
\begin{equation}\label{STRATELE-4}
f(x)= \left \{  \begin{array}{l}
f(0)e^{kx}, \hspace{2.5cm}  x \leq  0;  \\
f(0)e^{kx}, \hspace{2.5cm}  x \in [0,  L]; \\                                                   
f(0)e^{kL}e^{-k(x-L)}, \hspace{1cm}   x \geq L.
\end{array} \right.
\end{equation}
We use the notation $f_i=f(x_i), \,\, T_i=T(x_i)$. From  \eqref{STRATELE-3} 
and \eqref{STRATELE-4} we obtain  
\begin{equation}\label{S-6}
  (\mu_i-\mu_{i+1})f_i=  
\frac{ kU(\mu_{i+1}-\mu_i) - k^3 T_i}{\sigma_i}f_i, \quad 
0 \leq i  \leq N-1,
\end{equation}
\begin{equation}\label{S-7}
  (\mu_N +\mu_O)f_N=  
\frac{ kU(\mu_O-\mu_N) - k^3 T_N}{\sigma_N}f_N.                      
\end{equation}
From \eqref {STRATELE-A} we have  $(\mu_i-\mu_{i+1})<0$ for $0\leq i \leq N$,  
then \eqref{S-6} gives us 
\begin{equation}\label{S-9}
  \sigma_i \rightarrow \infty \mbox{  for  } 
 k \rightarrow  \infty,    \quad i=0,...,N-1. 
\end{equation}

Only $\sigma_N$ has 
a maximum  positive value with respect to the wave number $k$. 

If all  jumps $(\mu_i-\mu_{i+1})$ and  all surface tensions $T_i$ are equal, 
then $\sigma_0=\sigma_1=...= \sigma_{N-1} $ and we have only two different 
growth rates. 

The formula  \eqref{S-7}  is quite similar with the Saffman-Taylor value 
\eqref{SIGMA_ST01},   only  $\mu_W$  is replaced with $\mu_N$. 
\hfill   $\square$ 

\vspace{1cm}

{\it Remark 1.} Consider increasing  surface tensions $ T_i$ in \eqref{S-6}. 
Then for a fixed (large enough) $k$, the  eigenvalues will increase as well. 
This means that the  surface tensions amplify the instability. This is in 
contradiction with the  experimental results and also with the  Saffman-Taylor 
formula.

\hfill   $\square$ 

\vspace{0.5cm}

{\it Remark 2.}  The possible solutions of the  equation \eqref{ZT008} are 
$$ f(x) = A\exp(kx) + B \exp(-kx).                                           $$
It is possible to have $A=0, B=1$ and in this case we get
\begin{equation}\label{STRATELE-5}
f(x)= \left \{  \begin{array}{l}
f(0)e^{kx}, \hspace{3.2cm}  x \leq 0; \\  
f(0)e^{-kx}, \hspace{3cm}  x \in [0, L]; \\                                                  
f(0)e^{-kL}e^{-k(x-L)}, \hspace{1.5cm}   x \geq L.
\end{array} \right.
\end{equation}
Some elementary calculations show that the above eigenfunction   give us bounded
eigenvalues with respect to $k$. Moreover, the maxium (positive)  values of the 
eigenvalues can be arbitrary small if the viscosity jumps are small enough. 
It seems that in  \cite{DAR-DING}  it was omitted  the existence of the eigenfunction 
\eqref{STRATELE-4},  which is used in our paper. \hfill $\square$

\begin{center}

{\bf 4. The 3-layers Hele-Shaw flow with variable intermediate viscosity.  }

\end{center}

We consider $\mu_W < \mu_O$ and the  continuous linear intermediate viscosity 
\begin{equation}\label{continuu}
\mu(x) = (\mu_O-\mu_W)x/L + \mu_W,  \quad \forall x \in [0, L].
\end{equation}
We multiply with the {\it unknown} amplitude $f$ in  \eqref{ZT008}, we integrate 
on $[0,L]$ and get
\begin{equation}\label{PRE-SIGMAA}
 (\mu f_xf)^+(0)-(\mu f_xf)^-(L) + 
 \int_{0}^L \mu f_x^2 +     k^2 \int_{0}^L \mu f^2 = 
\frac{k^2}{\sigma} \int_{0}^L \mu_x f^2,    
 \end{equation}                      
where we use the notation $(FGH)(x):=F(x)G(x)H(x)$.

It is important to see that we have not jumps  of the viscosities in  
$x=0, x=L$. 
Then we consider, like in \cite{GOR-HOM-1}, that the surface tensions   
are equal to zero in these points and from the conditions \eqref{FAR-FIELD},  
\eqref{LAPLACE002}    we get 
$$ (\mu^+f_x^+)(0)= k \mu_Wf(0),   \quad (\mu^-f_x^-)(L)= - k \mu_Of(L). $$
Therefore the relation \eqref {PRE-SIGMAA}  lead  us to  the following growth 
rate, denoted by  $\sigma_L$: 
\begin{equation}\label{SIGMA-CONT}
 \sigma_L = \frac{ k^2  \int_{0}^L \mu_x f^2}
{\mu_Wk f^2(0) +  \mu_O kf^2(L) +  \int_{0}^L ( \mu f_x^2 + k^2 \mu f^2 )}.
\end{equation}
We neglect  some positive terms  in the denominator and obtain the estimate 
\begin{equation}\label{SIGMA-CONT-B}
 \sigma_L \leq \frac{ \int_{0}^L \mu_x f^2} {\int_{0}^L \mu f^2 }   \leq
\frac{ \mu_O-\mu_W}{L \mu_W}.
\end{equation}
In the  case  of a continuous (and {\it not} linear) viscosity $\mu$, we get 
\begin{equation}\label{SIGMA-CONT-B2}
\sigma_C \leq \frac{Max_x \,\, (\mu_x)}{ Min_x \,\, (\mu) }.
\end{equation}
Both  above estimates are not depending on $k$. Moreover, from \eqref{SIGMA-CONT-B}
it follows
\begin{equation}\label{SIGMA-CONT-B3}
 L \rightarrow \infty   \Rightarrow \sigma_L  \rightarrow 0.                
\end{equation}
On the page 3 of \cite{GEOLOGIC} is
considered a linear viscosity profile in a porous medium.

\begin{center}

 {\bf  5. Conclusions}

\end{center}

 The interface between two Newtonian immiscible fluids   in a rectangular Hele-Shaw cell is 
unstable  when the  displacing fluid is less viscous.

An intermediate fluid  with  a variable  viscosity  between the displacing fluids can 
minimize  the  Saffman-Taylor instability  -  see  \cite{CARASSO}, \cite{GILJE}, \cite{GOR-HOM-1}, 
 \cite{GOR-HOM-2}, \cite{SHAH}, \cite{SLOBOD}, \cite{UZOIGWE}.

The multi-layer Hele-Shaw model, consisting  of $N$ intermediate fluids with constant viscosities 
was studied  in  \cite{DAR-2}, \cite{DAR-DING}, \cite{D7}, \cite{D6}. The 
main result  of these papers was following: If  all surface tensions  verify some conditions  and  
$N$  is large enough,  then an arbitrary small (positive) upper bound of the growth rates can be 
obtained.

In this paper we point  out  a significant difference between  the displacement with constant 
intermediate  viscosities and the  displacement with a suitable variable intermediate viscosity. 
In the first case, if the viscosity-jumps are positive in the flow direction, then some eigenvalues 
become infinite for large wavenumbers - see   {\it Proposition 1}. 
In the second  case we can almost suppress  the  Saffman-Taylor instability - see the formula  
\eqref{SIGMA-CONT-B3}.

The growth constants given by the formulas \eqref{S-6}  tend to $\infty$ for large values
of $k$. Moreover, if the  surface tensions $T_i$  appearing in these formulas are very large, then  
the instability is amplified. This is in contradiction with the Saffman-Taylor stability
criterion.

Our main conclusion is following.  A possible strategy  to minimize the Saffman-Taylor  instability  is 
to use  the three-layer Hele-Shaw model  with a suitable {\it variable} intermediate  viscosity. On 
this   way, the instability of the two-layer flow can be almost suppressed.

Gelu I. Pa\c{s}a

Simion Stoilow Institute of Mathematics of Romanian  Academy

Calea Grivitei 21, Bucuresti Sector 1, Romania

e-mail: gelu.pasa@imar.ro 

\end{document}